\documentclass[10pt, conference]{IEEEtran}
\usepackage{amsmath,amssymb}
\usepackage{geometry}
\usepackage{graphicx}
\usepackage{url}
\usepackage{mdwlist}
\usepackage{color}
\usepackage{cite}
\usepackage{subfigure}

\newtheorem{theorem}{Theorem}[section]
\newtheorem{proposition}[theorem]{Proposition}

\newtheorem{example}{Example}

\newtheorem{definition}{Definition}

\newtheorem{remark}{Remark}

\begin{document}

\title{A cooperate-defect model for the spread of deviant behavior in social networks}
\author{Sarah Rajtmajer$^*$, Christopher Griffin$^{*\dagger}$, Derek Mikesell$^\dagger$, Anna Squicciarini$^+$\\ Department of Mathematics$^*$, College of Information Sciences and Technology$^+$\\Applied Research Laboratory$^\dagger$ \\ The Pennsylvania State University\\ University Park, PA, USA}
\maketitle
\begin{abstract}
We present a game-theoretic model for the spread of deviant behavior in online social networks. We utilize a two-strategy framework wherein each player's behavior is classified as normal or deviant and evolves according to the cooperate-defect payoff scheme of the classic prisoner's dilemma game. We demonstrate convergence of individual behavior over time to a final strategy vector and indicate counterexamples to this convergence outside the context of prisoner's dilemma. Theoretical results are validated on a real-world dataset collected from a popular online forum.
\end{abstract}

\section{Introduction}
Commenting systems on the Social Web have grown in popularity over the past few years, from blogs and social media sites like YouTube and Flickr to major news sites like NYTimes.com.

 Following this trend, episodes of abuse online are proliferating \cite{Chhabra:2011,Gao,cyberb,Lee:2010}.  Generally, any behavior that is "antisocial",  destructive, negative and offensive is considered as abusive, although abusive behavior comes in many forms ranging from minor to extremely harmful.  Some of such forms include grieving, trolling, flaming, harassment, threats trolling, multiple accounts, shared accounts, advertising, plagiarism etc. 

Users' reactions to antisocial peers may range drastically, based on their personality traits, predisposition toward certain behaviors, level of influence, and other contextual factors. Users may choose to engage in the discussion by mimicking the antisocial peer, confronting him and possibly reporting the behavior to superusers, simply ignoring the negative influence, or even leaving the network.
Over time, many popular real-world communities such as Reddit  \cite{Reddit}  have suffered damaging effects resulting from those few participants that choose to behave in a manner that is counter to established norms of behavior.

To date, despite a significant amount of work on the detection of social spammers, deception \cite{caspi}, collective attention spam \cite{Lee:2010,Thomas:2011,Kantchelian,Hu:2013,TGS:2012} and online vandalism, the dynamics underlying online abusive behavior remain uncertain \cite{sternberg2012}.

Toward developing a better understanding of this phenomenon, in this paper, we aim to build mathematical models to investigate the patterns of persuasion in online social networks, with emphasis on antisocial behavior.  We design an evolutionary game over a social network graph to describe how users interact with, influence and are influenced by both cooperative and antisocial behavior. During the game, members of the community may behave cooperatively, i.e. participating constructively in the network, or defectively, i.e. antisocial behavior. Each pairwise interaction between users yields a distinct payoff to each user, according to the extent of cooperation or defection exhibited by both in that interaction. At each iteration of game, every user observes his own total payoff as well as the actions and payoffs of his peers. We allow users to change their behavior accordingly, and we posit that this change in behavior will generally serve to mimic peers with observed higher payoff.

To the best of our knowledge, this is the first effort investigating this class of models to explain the spread of online abusive behavior.

We validate some of our findings on a real-world dataset collected from the web.  Precisely, we model the interactions amongst users who are part of a threaded online community and discuss topics of interest with other peers.  Each user post is assessed and labelled on a scale from most cooperative to most deviant.  The trajectories of these scores are then tracked within the context of the social network graph and compared against expected values of our theoretical evolutionary game.

Following we present a summary of relevant, related work in the literature, as it compares to our present study. We outline formally the problem we study and the game theoretic framework we have established, and prove some important theoretical properties of this game for both the special class of complete graphs and in the general case.  Finally, we compare our expectations with measured deviance on a real-world online dataset.

\section{Related Work}

Our work relates to the body of work on online deviance as it has been investigated by the computer science and by mathematical modeling approaches.

Several recent works focus on detecting social spammers \cite{Thomas:2011,Lee:2010,Kantchelian,Hu:2013,TGS:2012,java2006}. Social spammers, according to this body of work, are users controlled either by humans or bots, who use social networking sites and in particular their social connections to promote products, advertise events, or simply post useless and/or inappropriate comments.  Lee et al. \cite{Lee:2010} studied  social spammers in online social networks. To do so, they deployed social honeypots for harvesting deceptive spam profiles from social networking communities, and created spam classifiers using machine learning methods (e.g., SVM) based on a variety of features. Similarly, Kantchelian \cite{Kantchelian} developed an approach for detecting comment spam by leveraging the \textit{informativeness} level of a comment; he showed that spammers' comments have low information levels.  Hu et al. \cite{Hu:2013} also proposed a comprehensive framework for social spam detection, based on social network content, attaining a high accuracy of detection.
Our focus is not on social spammers only but rather on activities related to vandalism and misbehavior, which may or may not be generated by social spammers.

To deal with vandalism and bots, several tools exist (e.g. \cite{abuse,onlinetool}). Automated bots (e.g., Cluebot), filters (e.g., abusefilter), and editing assistants (e.g., Huggle and Twinkle) all aim to locate acts of vandalism. Such tools work via regular expressions and manually-authored rule sets. Though useful, these systems are unable to explain or predict where instances of abusive and deviant behaviors will occur, and are often limited to language detection.

Finally, our work parallels the body of work on free-riding in peer-to-peer systems \cite{feldman2006,thommes2006}. Peer-to-peer systems are designed to allow users to connect with others and share resources. Similar to online communities, users are free to access and contribute as much as desired, and few controls are in place.
As a result, in p2p systems peers may abuse their connections by exploiting other peers' resources, refusing to share owned resources, sharing broken or corrupted resources, etc., draining the network without contributing it.
In online communities,  the health of the community is heavily based on individual peers' reactions to selfish behavior, which they may choose to emulate or disengage from. Punishment mechanisms can also be put in place, although these are often considered not to be truly effective.
To tackle these issues, the most common solution is the implementation of incentive-based mechanisms. Incentives are applied in certain online forums, whereby end users are given special roles and privileges as a reward for good behavior.

From a theoretical perspective, our work can be placed in the context of the DeGroot model \cite{DeGro74} in which an individual changes her opinion dynamically and in part through imitation. In that model, a discrete time Markov chain forms the underlying behavior. From a game theoretic perspective, Morris \cite{Morris00} studies behavioral contagion in coordination games. Our work in this paper considers Prisoner's dilemma, but our main result (Proposition 5.1) is applicable to a broader class of games.  In addition to this, our work is inspired by work by the work of Jackson et al. \cite{jackson1996,jackson2003,jackson2008,jackson2005} when we consider the proposed model on the presence of network changes \cite{GMRS14}.

In this work, we propose a model for the spread of deviance in online social networks.  We do not propose mechanisms for punishment and reward of deviant behavior, but instead we model the inherent gain or loss for a user in the network who exhibits antisocial, deviant behavior, and accordingly, how this behavior as well as perceived gain or loss resulting from this behavior, effects the strategies of other users in the system over time. 

\section{Problem Statement}

Consider a network of users represented as a graph, wherein users are represented as nodes in the system and edges may represent stated friendships, recorded interactions or generally any indication that those two users have some influence on one another.  Each player is assigned an initial probability of choosing to cooperate during each interaction which we refer to as the player's initial \emph{strategy}.  In an online SN, the initial strategy represents individual predisposition for cooperation, separate from the effects of peer influence. \footnote{ In practice, initial strategies may be assigned to each user randomly, from a distribution of strategies over the entire population.  Or, if available, prior indicators of individual user behavior may be utilized to provide a better estimate.}

From this initial set of strategies, we introduce an evolutionary game wherein pairwise interactions amongst users yield a distinct payoff to each player, according to some payoff matrix. A player's total payoff at each iteration is given as the sum of his payoffs over all of his pairwise interactions. In our study, we adopt a prisoner's dilemma (PD)-type payoff matrix, wherein users may choose to cooperate or defect at each interaction and payoffs follow according to a canonical PD matrix. In the context of detecting deviant behavior online, the defector like the deviant user in a social network aims to influence or take advantage of a cooperative user and hence gains greatest benefit by an interaction with one such user. The cooperative user, on the other hand, suffers a negative payoff in this interaction,  while the cooperate-cooperate and defect-defect interactions are more neutral.

We assume that each player observes his own payoff and strategy at each iteration of the game, as well as the total payoff and the strategy for each of his neighbors. At each iteration of the game, a player may choose to change his strategy by mimicking successful strategies in his neighborhood, proportionally to their relative success.  This play/strategy revision procedure is repeated for some fixed number of iterations or until convergence (guaranteed, see below).  We assume a static graph in this work, and examine a dynamic graph in a follow-up study \cite{GMRS14}.

\section{Preliminaries}

Consider our model of behavioral evolution on a graph $G = (V,E)$ on $n$ nodes. Let $\mathbf{A} \in \mathbb{R}^{2 \times 2}$ be a payoff matrix. Following the prisoner's dilemma framework, we let \[\mathbf{A} = \left[ \begin{array}{cc}
a & b\\
c & d \end{array} \right]\]
where $c>a>d>0>b$.

If $i \in V$, let $x_i(t) \in [0,1]$ be the probability that player $i$ will play her first strategy at time (iteration) $t$. Thus, the strategy vector for Player $i$ is:
\begin{displaymath}
\mathbf{x}_i = \begin{bmatrix} x_i \\ 1 - x_i \end{bmatrix}
\end{displaymath}
The payoff to Player $i$ is:
\begin{equation*}
P_i = \sum_{j \in N(i)} \mathbf{x}_i^T\mathbf{A}\mathbf{x}_j
\end{equation*}
where $N(i)$ is the neighborhood of $i$ in $G$. In this paper, we will consider only two-strategy games; therefore let $\mathbf{x}$ be the vector of $x_i$ values. Define:
\begin{multline}
\kappa_{ij}(\mathbf{x}) := \\ \frac{\mathcal{H}\left(P_j(t) - P_i(t)\right)\left(P_j(t) - P_i(t)\right)}
{\sum_{k \in N(i)}\mathcal{H}\left(P_k(t) - P_i(t)\right)\left(P_k(t) - P_i(t)\right)}
\label{eqn:kappa}
\end{multline}
where $\mathcal{H}$ the Heaviside step function (defined as 0 at 0). In the case when $P_i(t) = P_j(t)$ for all $j \in N(i)$, let $\kappa_{ij}(\mathbf{x}) = 0$. Note that $\sum_{j}\kappa_{ij}(\mathbf{x}) = 1$, just in case $\kappa_{ij}(\mathbf{x}) \neq 0$ for all $j \in N(i)$. Let:
\begin{equation}
f_i(\mathbf{x}) = \sum_{j \in N(i)} \kappa_{ij}(t)(x_j(t) - x_i(t))
\end{equation}
The strategy update rule is given by
\begin{equation}
x_i(t+\epsilon) = x_i(t) + \epsilon f_i(\mathbf{x})
\label{eqn:Imitation}
\end{equation}
Here: $\epsilon > 0$ and $\epsilon \ll 1$. This rule is a \textit{proportional success mimicking rule} in which players will drift toward (imitate) successful behaviors.


\section{Convergence of Players' Strategies to a Final Strategy Vector}

We examine the trajectory of individual users' strategies, assuming no changes in network structure. The addition or deletion of network ties being more infrequent than small shifts in individual behavior \cite{hoadley,Gross:2005}, we here capture those behavioral changes that precede a severing or creation of a structural tie. We claim that in this case, assuming a static network structure, there can be stable strategy equilibria. This holds for all network graphs, and we show that in the special case of the complete graph this vector has the form \[\mathbf{x_f}=[x_{1_f},x_{2_f},\ldots,x_{n_f}]\] where $x_{1_f}=x_{2_f}=\ldots=x_{n_f}$.

\subsection{Equilibrium Points of Player Strategies on an Arbitrary Network}
To study the fixed points of player strategies under the dynamics of Expression \ref{eqn:Imitation} it is easier to pass to the continuous dynamics:
\begin{equation}
\dot{x}_i = f_i(\mathbf{x}) = \sum_{j\in N(i)} \kappa_{ij}(\mathbf{x})(x_j - x_i)
\label{eqn:Dynamics}
\end{equation}
There are three types of equilibria for the differential system defined by Equation \ref{eqn:Dynamics}:
\begin{definition} A \textit{Type 1 Equilibrium} occurs when $x_i = x_j$ for all $i,j$.
\textit{A Type 2 Equilibrium} occurs when $P_{i}(\mathbf{x}) = P_{j}(\mathbf{x})$ for all $i,j$ and there is at least one pair $i,j$ so that $x_i \neq x_j$. A \textit{Type 3 Equilibrium} is any equilibrium not satisfying the conditions of Types 1 or 2.
\end{definition}

Note, a Type 3 equilibria may contain groups (e.g., cliques) of vertices that all share a strategy within group but have different strategies among the groups. It is clear that every system of the type given by Equation \ref{eqn:Dynamics} is degenerate in the sense that it has an infinite number of equilibria. In this case, it is similar to the SIR dynamics discussed in \cite{Hethcote00themathematics}. We show that under certain conditions, Type 1 equilibria may be (at worst) neutrally stable. Note first:
\begin{enumerate*}
\item if $p = i$:
\begin{equation}
\frac{\partial f_i}{\partial x_p} = \sum_{j \in N(i)}\left\{\frac{\partial \kappa_{ij}(\mathbf{x})}{\partial x_p}(x_j - x_i) - \kappa_{ij}(\mathbf{x})\right\}
\end{equation}
\item if $p \in N(i)$:
\begin{equation}
\frac{\partial f_i}{\partial x_p} = \sum_{j \in N(i)}\left\{\frac{\partial \kappa_{ij}(\mathbf{x})}{\partial x_p}(x_j - x_i)\right\} + \kappa_{ij}(\mathbf{x})
\end{equation}
\item Otherwise:
\begin{equation}
\frac{\partial f_i}{\partial x_p} = \sum_{j \in N(i)}\frac{\partial \kappa_{ij}(\mathbf{x})}{\partial x_p}(x_j - x_i)
\end{equation}
\end{enumerate*}
We are liberally abusing the derivative operator here, since $\kappa_{ij}(\mathbf{x})$ is clearly not smooth. When we refer to its derivative we refer to its derivative as a generalized function with appropriate Dirac delta distributions used.

\begin{proposition} If $\mathbf{x}^*$ is an equilibrium point of Type 1 and $\sum_{j \in N(i)}\kappa_{ij}(\mathbf{x}^*) > 0$ for at least one $i$, then $\mathbf{x}^*$ is at worst neutrally stable.
\label{prop:Prop1}
\end{proposition}
\begin{IEEEproof} We make use of the face that $x\delta(x) = 0$ in the generalized density product, where $\delta(x)$ is the Dirac delta. From our observations and the fact that $x_i = x_j$ for all $i,j$, we have that:
\begin{equation}
\left.\frac{\partial f_i}{\partial x_p}\right\rvert_{x = \mathbf{x}^*} = \begin{cases}
 - \sum_{j \in N(i)} \kappa_{ij}(\mathbf{x}^*) & \text{if $p = i$}\\
 \kappa_{ij}(\mathbf{x}^*) & \text{if $p \in N(i)$}\\
0 & \text{otherwise}
\end{cases}
\end{equation}
This implies:
\begin{equation}
\left.\frac{\partial f_i}{\partial x_p}\right\rvert_{x = \mathbf{x}^*} = \begin{cases}
 - 1 & \text{if $p = i$}\\
 \kappa_{ij}(\mathbf{x}^*) & \text{if $p \in N(i)$}\\
0 & \text{otherwise}
\end{cases}
\end{equation}
Applying the Gershgorin Disk Theorem, we know that every eigenvalue of the Jacobian lies within a disk in the complex plane centered at $-1$ with radius $1$. Thus, the real-parts of these eigenvalues are non-positive and the resulting fixed point must be neutrally stable.
\end{IEEEproof}

\begin{remark} It follows from Equation \ref{eqn:kappa}, that at least one row of the Jacobian matrix will contain all zeros, since there is at least one vertex with greatest payoff. Thus, at least one eigenvalue must be identically zero and the resulting Type 1 fixed points are never asymptotically stable; however they do exhibit a basin of attraction as we illustrate in two examples.
\end{remark} 

\begin{example} Consider the randomly generated graph with 10 vertices in Figure \ref{fig:RandomGraph} (a). We use the prisoner's dilemma payoff matrix:
\begin{equation}
\mathbf{A} = \begin{bmatrix}3 & -7\\5 & 2\end{bmatrix}
\end{equation}
\begin{figure}[htbp]
\centering
\subfigure[]{\includegraphics[scale=0.18]{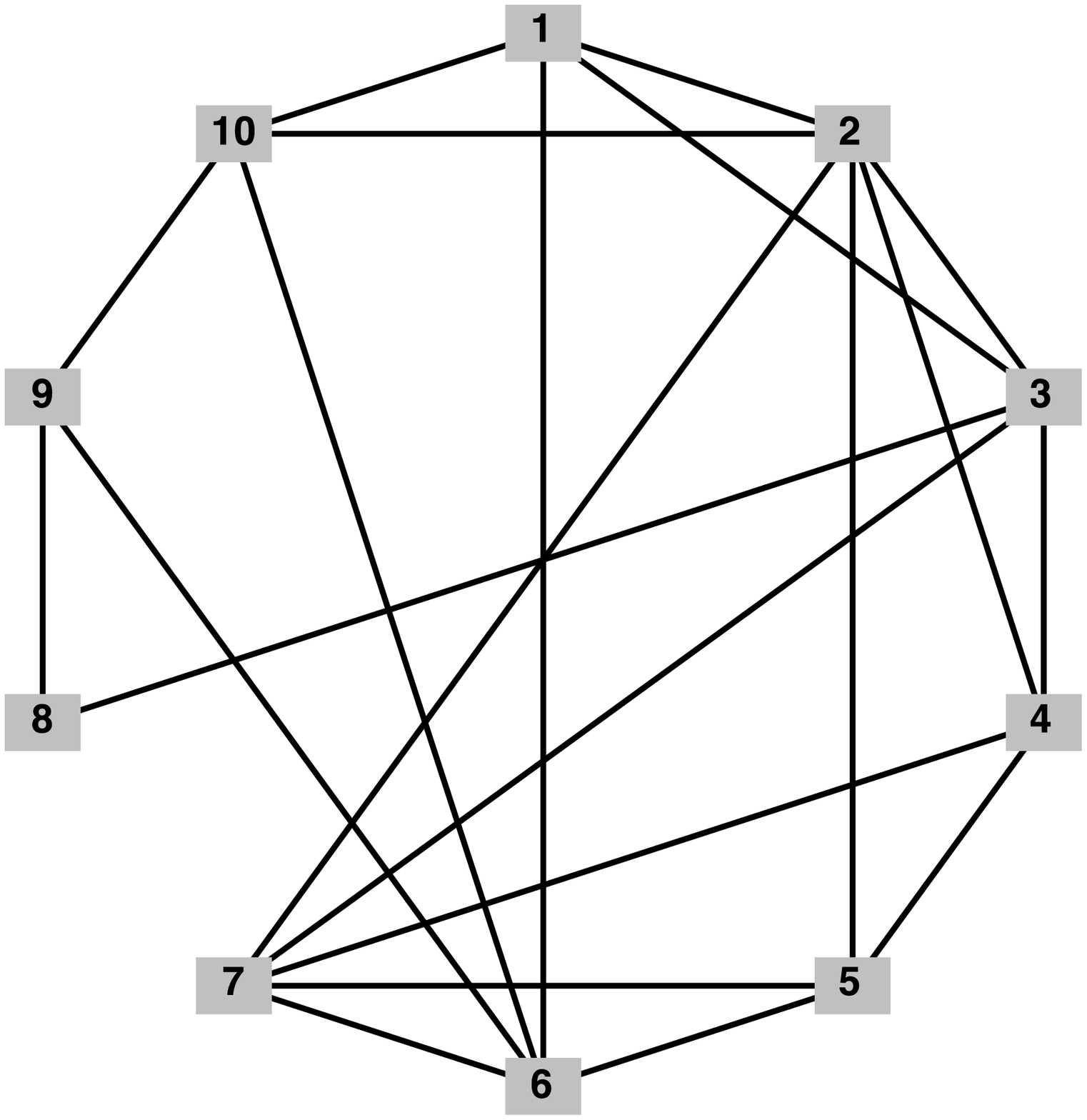}}
\subfigure[]{\includegraphics[scale=0.18]{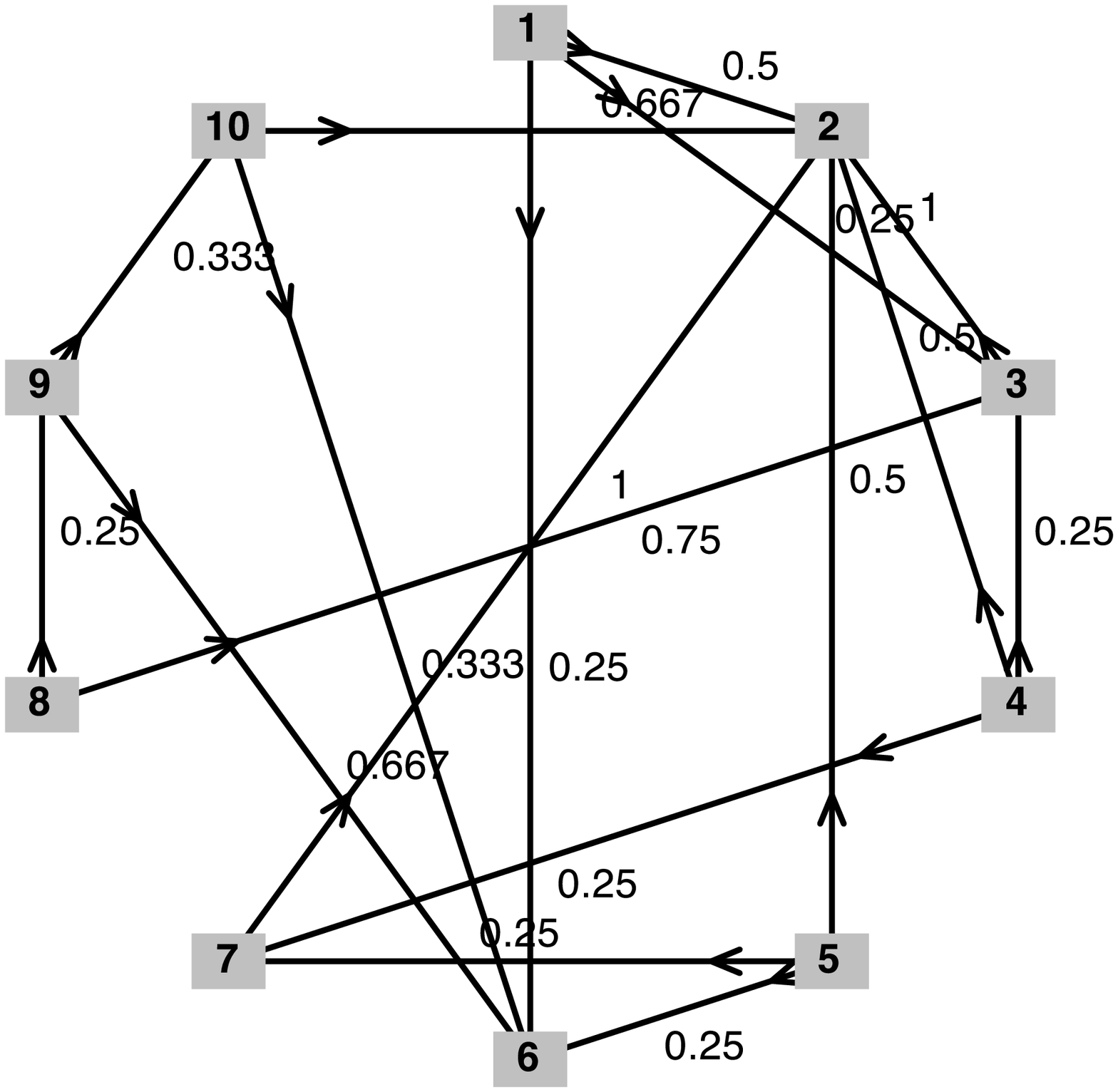}}
\caption{(a)A random graph on 10 vertices used to illustrate the neutral stability of Type 1 fixed points. (b) The weighted directed \textit{imitation graph} at the fixed point.}
\label{fig:RandomGraph}
\end{figure}
Any non-extreme (i.e., not all cooperating or defecting) Type 1 fixed point will suffice for illustrate the stability properties of these fixed points. Without loss of generality, will use the $x^*_i = 0.795$ for $i = 1,\dots,10$. If we compute $\kappa_{ij}(\mathbf{x}^*)$ for each pair, we can construct the weighted directed \textit{imitation graph}, shown in Figure \ref{fig:RandomGraph} (b). Note, the negative of the Laplacian for the imitation graph is the Jacobian matrix of the dynamical system (when we extend the notion of the Laplacian to weighted directed graphs). Vertices with $0$ out-degree have highest local payoff and indicate directions in which we may perturb the equilibria and remain in the basin of attraction. By way of example, consider the new point strategy set:
\begin{multline}
\mathbf{x}' = [ 0.79, 0.795, 0.79, 0.8, 0.795, 0.795,\\ 0.792, 0.795, 0.795, 0.7]
\end{multline}
Here we have perturbed players 1, 3, 4, 7, and 10 from their equilibrium position, leaving players 2, 5, 6, 8 and 9 at the original equilibrium. Analysis of the imitation graph suggests the critical players are 2 and 6, both of whom have $0$ out-degree (and thus zero rows in the Jacobian matrix). The return to equilibrium is illustrated in Figure \ref{fig:Equilibrium} (a). This does not mean that the basin of attraction for this point is a ball. From the the non-equilibrium point:
\begin{multline}
\mathbf{x}' = [ 0.795, 0.795, 0.795, 0.795, 0.795, 0.79,\\ 0.795, 0.795, 0.795, 0.795]
\end{multline}
the system converges to a nearby Type 3 equilibrium, illustrating the neutral stability of the original Type 1 point.
\begin{figure}[htbp]
\centering
\subfigure[]{\includegraphics[scale=0.18]{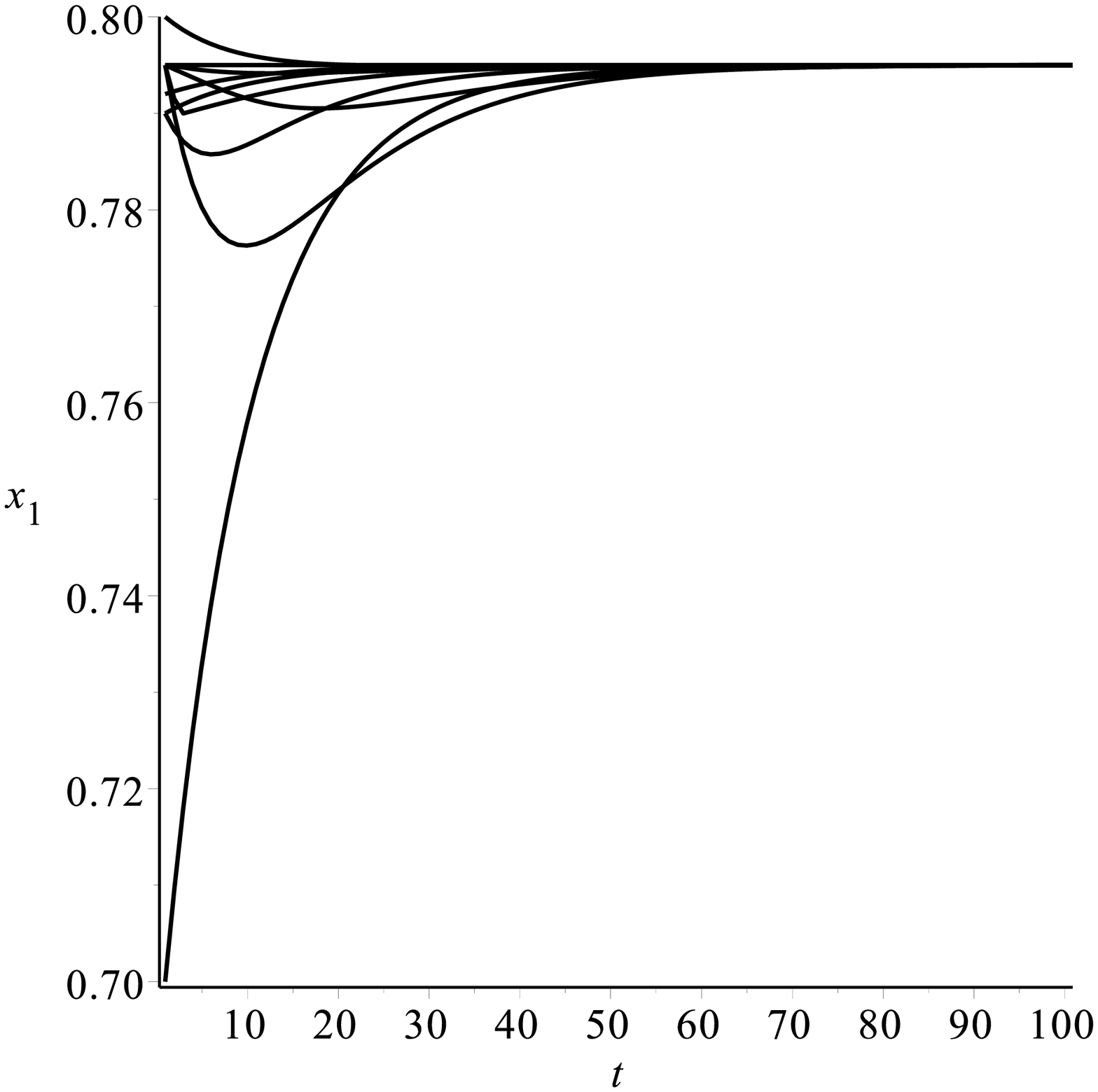}}
\subfigure[]{\includegraphics[scale=0.18]{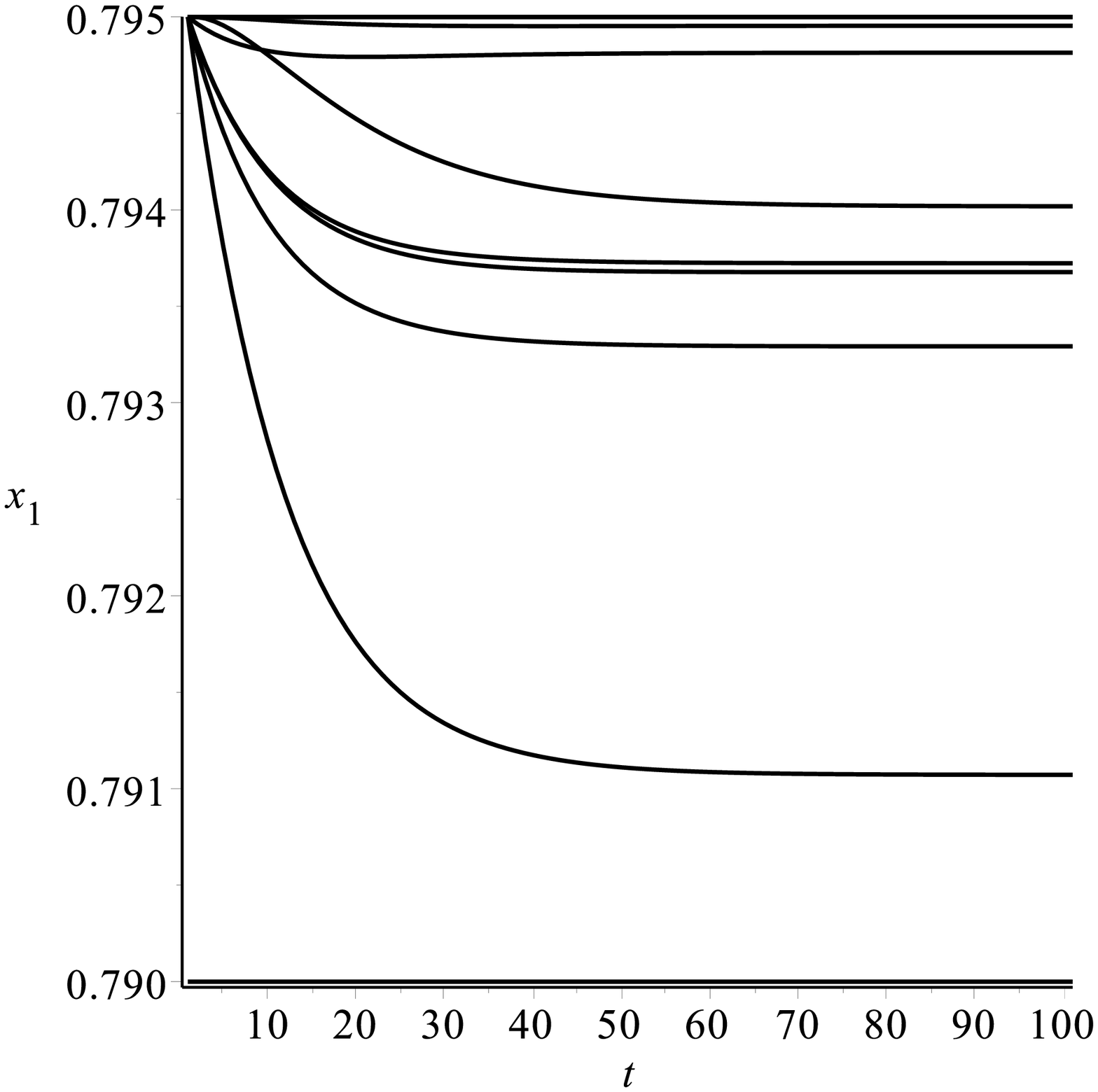}}
\caption{(a) The system is shown returning to equilibrium after being perturbed. (b) Displacing Player 2 from equilibrium causes the system to move to a new Type 3 equilibrium.}
\label{fig:Equilibrium}
\end{figure}
An interesting element of this example is Player 10, who becomes significantly more defective is made more cooperative by peer imitation.
\end{example}

\begin{remark} Notice the previous results do not rely on a Prisoner's dilemma assumption and the previous proof should generalize to payoff matrices in $\mathbb{R}^{n \times n}$ with appropriate changes made.
\end{remark}

\begin{remark} Notice this analysis says nothing about Type 1 equilibria that also satisfy the criteria of Type 2 equilibria. In this case, the Jacobian matrix consists of all zeros and thus the imitation graph has no edges.
\end{remark}

\begin{remark}
Analysis of Type 2 and 3 equilibria is even less satisfying. In the case of Type 2 equilibria, we see we are at a point where all derivatives of $f_i(\mathbf{x})$ fail to converge and consequently nothing can be said about these points, however the stability is highly questionable. For a Type 3 equilibrium point satisfying the assumptions of Proposition \ref{prop:Prop1}, the Gershgorin disk theorem gives no information about the convergence other than that these points may be stable or unstable depending on the nature of the graph, the point and the payoff matrix. In practice, we see convergence to Type 3 equilibria in which vertices can be partitioned into a few strategic species; i.e., Type 3 equilibria, in practice, look like the amalgamations of several Type 1 equilibria. Note, Type 3 equilibria satisfying the definition of Type 2 equilibria may also have unusual stability properties.
\end{remark}

\subsection{Convergence of Player Strategies in the Complete Case}

In the case of the complete graph, we claim that all players will converge on a single strategy (Type I), in particular the initial strategy of the most deviant user in the graph. While the complete graph has a very strict tie structure and online networks are generally not complete, this result provides insight into local behavior in large networks since their typical small-world structure is highly clustered \cite{Watts-Colective-1998}, with densely-linked clusters also tending to be more homogenous \cite{homophily}.

\begin{theorem}
In the absence of network changes, the strategies of all players in a completely connected network will converge to a unique value, in particular to the minimal strategy (least cooperative) in the initial strategy vector.
\end{theorem}

\begin{IEEEproof} Without loss of generality, let $x_l = x_1(t_0) \leq x_2(t_0) \leq \ldots \leq x_n(t_0)$ be the initial strategies of $n$ completely connected nodes.

The initial payoffs to each are node are ordered \[P_1(t_0) \geq P_2(t_0) \geq \ldots \geq P_n(t_0)\]
since according to our model, in each pairwise interaction the node less likely to cooperate has a higher payoff.

Since each player $i$ observes all other players in the complete graph, he will update his strategy according to:
\begin{multline} \label{updatecomplete}
x_i(t+1)=x_i(t)+ \\
\epsilon \cdot \sum_{j\in J }{\frac{(P_j(t)-P_i(t))(x_j(t)-x_i(t))}{\sum_{j\in J}{P_j(t)-P_i(t)}}}
\end{multline}
where $J$ is the set of nodes such that $x_j(t)<x_i(t)$. This is a natural modification of Equation \ref{eqn:Imitation}. Thus we see that, $x_i(t+1)<x_i(t)$ for all $x_i$ with the exception of $x_l$ which remains unchanged. Thus we have
\begin{equation}
\lim_{t \to \infty}x_i(t)=x_l \quad \forall v_i\in V.
\end{equation}

\end{IEEEproof}

\section{The Spread of Deviance on BodyBuilding.com}
In this section, we describe our empirical measurements of the models against a real-world dataset collected from the web.

\subsection{Data Collection and Measures for Deviance}
We collect data from a real-world online bulletin system,  known for  the controversial and active nature of its discussions  (BodyBuilding.com, a popular and longstanding site for body builders and fans).   Bodybuilding.com is powered on a Vbulletin portal, wherein discussions are grouped by topic, and each discussion is further given by a collection of threads. Logged users can create a new thread and respond to any contributed comment within a thread.
For the purpose of our validation, we specifically concentrate on a network of users active on the ``Religion/Politics'' sub-forum, which collects several heated discussions on controversial topics. We build the users' network by joining users who have commented on the same thread with an undirected edge, under the assumption that users who post to a thread have some degree of familiarity with the other comments posted. This assumption is supported by the longstanding presence of many of the community members, who contribute regularly to the same topics.

We collected a total of $62,060$ posts, over $787$ threads, obtaining a network of $2586$ distinct nodes (users) joined by $62,931$ unweighted, undirected edges.

\begin{figure}[h!]
\centering
\includegraphics[width=0.4\textwidth]{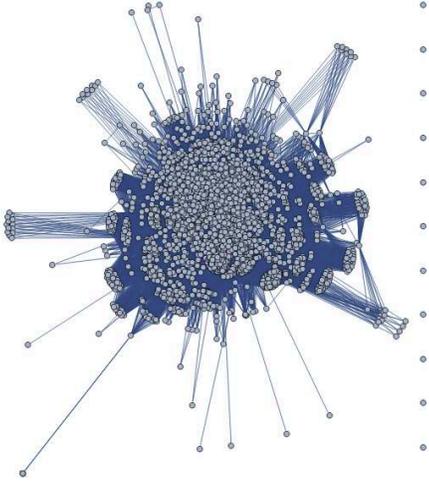}
\caption{Interactions amongst users posting to the "Religion/Politics" forum on www.bodybuilding.com.}
\label{fig2}
\end{figure}

In order to explore the behavior of users in this network within the theoretical game framework we have established, we split the network into two pieces by time. Particularly, we create an early network and a late network of user interactions obtained from the first $31,030$ posts and the second $31,030$ posts, respectively. This allows us to fit our models in two distinct temporal frames, comparing the evolution of individual deviant behavior.

  Each post is assigned a deviance score in $[0,1]$, taken as a weighted average of three measures: {\em language, sentiment} and {\em content relevance}.  Here by language, we mean the use of abusive, vulgar or inflammatory words.  To obtain a  measure for each post with respect to this dimension, we extract distinctive words from the text and compared them against dictionary of known abusive words,  to determine the ratio of abusive words used. If the result is above a  predefined threshold (e.g. $20\%$ in our context), we label the comment negative otherwise it is neutral.
	To determine the post's sentiment, we rely on the  AlchemyAPI tool \cite{alchemyapi}. Alchemy is a text mining platform with advanced natural language processing capabilities for semantic analysis. It is  widely used  both  in academia and  industry. In this regard, it is reasonable to assume a reliable accuracy in terms of sentiment classification. Through Alchemy, the sentiment of each post is marked as "positive", "negative" or "neutral" according  To confirm this hypothesis, we additionally validated  sentiment scores using the SentiStrength algorithm \cite{sentistrength}, which returned the same sentiment results for 87\% of the posts.
	
	Content relevance measures the extent to which a particular comment is degenerated relative to the post originating the discussion. It is measured by considering the mutual information (MI) of the comment,
with respect to the category or topic of the thread. For a given thread $T$ and a comment $W$, it is computed by measuring the
    the amount of information each term $w$ in  $W$ relates with the thread $T$, wherein the relationship strength (i.e. amount of information) is determined using the Wordnet dictionary as a reference thesaurus.
     The less cohesive the comment is with respect to the whole
thread, the more degenerated or out of context it is
likely to be.

The deviance scores for the subset of users present in both the early and late networks ($586$ total) were compared against expected results using the prisoners dilemma model for the spread of deviant behavior in social networks presented above. The early deviance score for each user was set as the initial strategy, while the late score was compared with the final strategy, obtained after running the PD simulation until convergence ($\epsilon f_i(\mathbf{x}) \leq$ for all $i$). This is consistent with sociological evidence that individuals will settle into behaviors over time and we have ensured convergence within our model.

\subsection{Empirical Analysis and Observations}
\begin{figure}[htbp]
\centering
\includegraphics[scale=0.4]{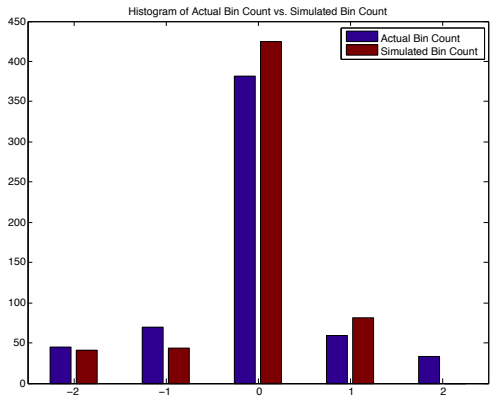}
\caption{A histogram showing the true vs. expected distribution of users into deviance bins.}
\label{fig:Histogram}
\end{figure}

Initial work with the data showed that fitting precise deviance values was impossible (and ill-conceived, since we were using a computerized method to determine deviance, which could add substantial noise to the measured values). Let $Y_0$ be the early deviance scores and $Y$ be the late deviance scores and let $S$ be the standard deviation of the $Y_0$. We binned the elements of $Y$ using $S$ and $Y_0$ so that if $Y_i \in Y_{0_i} \pm S$, then it was assigned Bin $0$. In general if:
\begin{equation}
Y_i \in Y_{0_i} \pm kS \text{ and } Y_i \not\in Y_{0_i} \pm (k-1) S,
\end{equation}
then $Y_i$ is in bin $\pm k$. For simplicity, we used five bins, labeled $-2$ through $2$. Letting $\hat{Y}$ be the estimated deviance values, we were also able to compute bin values using the same technique. The objective was to determine the general ability of the model to measure an individual's propensity to become more or less deviant as a function of time. The values of the payoff matrix $\mathbf{A}$ were fit to minimize the sum of square bin-error. The best-fit matrix is given by:
\begin{equation}
\hat{\mathbf{A}} = \begin{bmatrix} 0.1985 & -0.6989\\0.4927 & 0.0001\end{bmatrix}
\end{equation}
The confusion matrix that results from this model is given by:
\begin{displaymath}
\mathbf{C} = \begin{bmatrix}
28 & 11 & 6 & 0 & 0 \\
6 & 17 & 47 & 0 & 0 \\
7 & 14 & 319 & 42 & 0 \\
0 & 1 & 36 & 22 & 0 \\
0 & 0 & 16 & 17 & 0
\end{bmatrix}
\end{displaymath}
Rows are true bins, while columns are expected bins. From this we can compute the confusion matrix in probabilities:
\begin{displaymath}
\mathbf{P} = \begin{bmatrix}
0.62222 & 0.24444 & 0.13333 & 0 & 0 \\
0.085714 & 0.24286 & 0.67143 & 0 & 0 \\
0.018325 & 0.036649 & 0.83508 & 0.10995 & 0 \\
0 & 0.016949 & 0.61017 & 0.37288 & 0 \\
0 & 0 & 0.48485 & 0.51515 & 0
\end{bmatrix}
\end{displaymath}
Figure \ref{fig:Histogram} also summarizes our fit using a histogram.
This matrix is instructive on the capabilities and limitations of the modeling approach. We make the following observations:
\begin{enumerate*}
\item The model is very good at estimating when individuals will remain at approximately the same deviance level. There is an 83\% chance of correct classification in this case.
\item The model is reasonably good at estimating when individuals will become much more deviant (most likely because of the attractiveness of the deviant strategy in the prisoner's dilemma model).
\item The model tends to under-classify minor deviance and significantly under-classify individuals who move toward extreme non-deviance (i.e., who move to a +2 bin).
\end{enumerate*}
What is surprising is the model's tendency to under-represent minor shifts toward more deviant behavior, since the prisoner's dilemma has the strict Nash equilibrium defect. This suggests that this approach is promising as a way of modeling the emergence of more cooperative behaviors within networks. It also suggests (given the under-representation of minor deviance) that this game is not the ideal model for handling deviance. Further research in both these areas is required.

\section{Conclusion}

In presenting a general evolutionary game model for the spread of deviant behavior in social network graphs, we lay the foundation for an extension of the general contagion model for influence online to include the notion of individual payoff. We utilize a two-strategy prisoner's dilemma model here, but alternate models may be more deeply investigated in future work. Additionally, we assume a single payoff matrix across all users. Further research may incorporate multiple payoff matrices where this information is available or detect users whose anomalous actions indicate that they are operating with a different payoff scheme.  Validation of this theoretical framework with a large, online dataset gives preliminary indication that the model may indeed work well when tuned within a particular domain.

\section*{Acknowledgements}
Portions of Dr. Griffin's work were supported by the Army Research Office under Grant W911NF-11-1-0487. Portions of Dr. Griffin's, Dr. Squicciarini's and Dr. Ratjmajer's work were supported by the Army Research Office under grant W911NF-13-1-0271.

\bibliographystyle{plain}
\bibliography{BibDeviantBehavior}

\end{document}